\documentclass[12pt]{iopart}
\usepackage[dvips]{graphicx}
\usepackage{color,ulem}

\usepackage{iopams}  
\newcommand{\ignore}[1]{}
\newcommand{\beq}{\begin{equation}}
\newcommand{\eeq}{\end{equation}}
\newcommand{\mbold}[1]{\mbox{\boldmath $ #1 $}}

\begin{document}

\title{Photon and spin dependence of the resonance lines shape in the strong coupling regime}

\author{Seiji Miyashita}
\address{Department of Physics, Graduate School of Science, The University of Tokyo, 7-3-1 Hongo, Bunkyo-ku, Tokyo 113-8656, Japan}
\address{CRESTO, JST, 4-1-8 Honcho Kawaguchi, Saitama 332-0012, Japan}
\ead{miya@spin.phys.s.u-tokyo.ac.jp}

\author{Tatsuhiko Shirai}
\address{Department of Physics, Graduate School of Science, The University of Tokyo, 7-3-1 Hongo, Bunkyo-ku, Tokyo 113-8656, Japan}
\address{CRESTO, JST, 4-1-8 Honcho Kawaguchi, Saitama 332-0012, Japan}
\ead{shirai@spin.phys.s.u-tokyo.ac.jp}

\author{Takashi Mori}
\address{Department of Physics, Graduate School of Science, The University of Tokyo, 7-3-1 Hongo, Bunkyo-ku, Tokyo 113-8656, Japan}
\address{CRESTO, JST, 4-1-8 Honcho Kawaguchi, Saitama 332-0012, Japan}
\ead{mori@spin.phys.s.u-tokyo.ac.jp}

\author{Hans De Raedt}
\address{Department of Applied Physics, Zernike Institute for Advanced Materials, University of Groningen, Nijenborgh 4, NL-9747 AG Groningen, The Netherlands}
\ead{h.a.de.raedt@rug.nl}

\author{Sylvain Bertaina}

\address{IM2NP-CNRS (UMR 7334) and Universit\'{e} Aix-Marseille, Facult\'{e} des Sciences et Techniques, Avenue Escadrille Normandie Niemen - Case 142, FR-13397 Marseille Cedex, France}
\address{Department of Physics and The National High Magnetic Field Laboratory, Florida State University, Tallahassee, Florida 32310, USA}
\ead{sylvain.bertaina@im2np.fr}

\author{Irinel Chiorescu}
\address{Department of Physics and The National High Magnetic Field Laboratory, Florida State University, Tallahassee, Florida 32310, USA}
\ead{ic@magnet.fsu.edu}

\begin{abstract}
We study the quantum dynamics of a spin ensemble coupled to cavity photons. 
Recently, related experimental results have been reported, showing the existence of the strong coupling regime in such systems. We study the eigenenergy distribution of the multi-spin system (following the Tavis-Cummings model) which shows a peculiar structure as a function of the number of cavity photons and of spins. We study how this structure causes changes in the spectrum of the admittance in the linear response theory, and also the frequency dependence of the excited quantities in the stationary state under a probing field. In particular, we investigate how the structure of the higher excited energy levels changes the spectrum from a double-peak structure (the so-called vacuum field Rabi splitting) to a single peak structure.
We also point out that the spin dynamics in the region of the double-peak structure corresponds to recent experiments using cavity ringing while in region of the single peak structure, it corresponds to the coherent Rabi oscillation in a driving electromagnetic filed. Using a standard Lindblad type mechanism, we study the effect of dissipations on the line width and separation in the computed spectra. In particular, we study the relaxation of the total spin in the general case of a spin ensemble in which the total spin of the system is not specified. The theoretical results are correlated with experimental evidence of the strong coupling regime, achieved with a spin 1/2 ensemble.\\

\end{abstract}

\maketitle

\section{Introduction}
Quantum mechanics principles provides the fundamental basis for the development of future technologies which will integrate quantum processors with quantum memories. Electromagnetic impulses, like photons, can provide the link between the processing of quantum algorithms and temporary storage between computation steps. Namely, quantum information can be transferred by photons into a solid state memory able to preserve the quantum aspects of the information for a certain time. Initially, the coupling between atoms and resonators was first studied in atomic physics, since in the absence of significant atom-atom interaction, the atomic levels have large lifetimes compared to their solid-state counterparts. In such situation, the transfer of quanta between atoms and resonators can be observed.

The coupling between atoms and a standing electromagnetic wave was introduced by Tavis and Cummings \cite{JC,TC} in the 1960s, and calculated to increase with the square root of the number of atoms. This phenomenon has been observed for an ensemble of atoms \cite{Kaluzny_PRL83,Mondragon_PRL83} and single atoms \cite{Thompson_PRL92,Brune_PRL96} in cavity quantum electrodynamics experiments (cQED), in which photons and the atoms interact in a cavity. In more recent years, the use of a superconducting resonator with trapped molecules above it, has been proposed \cite{Rabl_PRL06} to implement cQED experiments on a chip. These initial studies made use of the electric-field component of an electromagnetic field to coherently exchange photons between the field and the system under study.

In solid state systems, magnetic two-level systems (like up and down spins) can have significantly longer coherence times compared to electrical analogues. It is therefore desirable to implement the cQED techniques developed for electrical coupling to magnetic coupling, although the magnetic-field component of an electromagnetic field offers much less coupling. This aspect is important for the implementation of quantum computing on a chip, using solid state materials, in which quantum information can be rapidly lost to the neighboring environment. Dipolar interactions between spins in such systems are a major cause of information loss, since it affects the spin coherence lifetime. But these processes can be reduced by diluting the spins in the host solid. Magnetic coupling between the electromagnetic field in the cavity and a spin ensemble has been demonstrated theoretically \cite{Imamoglu_PRL09} and experimentally \cite{Irinel,KuboY,Schuster,Amsuss_PRL11}, in the regime called strong coupling. This case requires a coupling strength larger than both the cavity's photon decay rate and the rate at which the spin losses its quantum state information. Aside coupling to a cavity, recent studies \cite{Semba_Nature11,KuboY_PRL11} show that diluted spin systems can be coupled with and exchange information with superconducting qubits, making them suitable quantum memories. 

The spin-photon coupling can be modeled by the Jaynes-Cummings (for one spin) or Tavis-Cummings model (for multi-spin), and it has been shown that the photon absorption or emission spectrum has a pair of peaks, called "vacuum-field Rabi splitting" \cite{Agarwal}. We study how the energy structure changes with the number of photons in the cavity, and investigate how the change causes the spectrum of the admittance in the linear response theory, and also the frequency dependence of the excited quantities in the stationary state under a probing field. 

When the number of photons is much smaller than the number of spins, the spectrum
shows the quantum mechanical double peak structure, while it shows a single peak corresponding to the magnetic resonance in an driving oscillatory field when the number of photons is much larger than the number of spins.

We also point out that the spin dynamics in the region of the double-peak structure corresponds to our experiments using cavity ringing (see also \cite{Irinel}) showing vacuum-field Rabi splitting. On the other hand, spin-photon system can show coherent Rabi oscillation, representing spin dynamics under a driving classical electromagnetic filed, in the region of the single peak structure. Between these two regions, namely when the number of spins and the number of photon in the cavity are similar, a chaotic spin dynamics appears. 

We also study line shape of the spectrum of photon absorption and emission in dissipative environments. We point out that the line shape is affected by several conditions: how many spins are involved, how strongly the system is driven, how strong the dissipations are, what types of  dissipation exist, and whether we consider the linear spectrum or the spectrum for nonlinear steady state under driving force.

\section{Model}
\subsection{Hamiltonian}

As system Hamiltonian we adopt the Tavis-Cummings model (a generalized Jaynes-Cumming model) ${\cal H}_0$ with a driving force $\xi$ \cite{JC,TC,Bishop}:
\beq
{\cal H}={\cal H}_0+{\cal H}_{\xi}
\eeq
where
\beq
{\cal H}_0=
\omega_0a^{\dagger}a+\omega_{\rm s}(\sum_i^N S_i^z + N/2) 
-g\left(a^{\dagger}\sum_i^NS_i^- + a\sum_i^NS_i^+ \right)
\eeq
\beq
{\cal H}_{\xi}=\xi\left(a^{\dagger}e^{-i\omega t} + ae^{i\omega t} \right).
\label{ham}
\eeq 
Here $\{\mbold{S}_i\}$ denotes a set of two-level systems, and we call them `spin' hereafter. Each is represented by the operator for a spin of $S=1/2$. Here, $\omega_0$ is the eigenfrequency of the cavity, $\omega_{\rm s}$ is the energy gap of a two-level system, and $\omega$ is the frequency of the driving force. Throughout the paper, we put $\hbar=1$. Here the system contains $N$ two-level systems. 
The dimension of the Hilbert space of the spin system is $2^N$. First, we consider the case that only the uniform mode (i.e., the uniform sum of spins of the system),
\beq
\mbold{S}=\sum_{i=1}^N \mbold{S}_i
\eeq 
couples to the cavity, and thus the total spin of the system is conserved. (We will study the case in which the total spin can be changed in the section V.) Therefore, the Hamiltonian is block-diagonalized with the total spin. The total spin is $S=N/2$ when the number of spins is $N$. In the system, the sum of the photon number $n$ and the magnetization, $M=\sum_{i=1}^N S_i^z$
\beq
C=n+M
\eeq
is conserved. Thus, the Hamiltonian is further block-diagonalized with this quantity.
To characterize the block, we introduce a quantity $n_{\rm max}=C+S$ which is the number of photons when all the spins are down, i.e., the state of $M=-S$. The size of the sub-block increases from 1 to $N+1$ where $n_{\rm max}$ is equal to $2S+1(=N+1)$.
For larger photon numbers, the size of sub-matrix is $N+1$, where extra photons exist even when all the spins are up (i.e., $M=S$). 

We adopt the basis in the form $|n,M\rangle$. The matrix elements are given by the relations
\beq
S^+|n,M\rangle=C_{S,M}|n,M+1\rangle, \quad C_{S,M}=\sqrt{S(S+1)-M(M+1)},
\eeq
and
\beq
a^{\dagger}|n,M\rangle=\sqrt{n+1}|n+1,M\rangle.
\eeq
The matrix of ${\cal H}_0$ is given as
\beq \hspace*{-23mm}
{\cal H}_0=
\tiny
\left(\begin{array}{rrrrrrrrrr}
0 & 0          & \cdots      &           &                              &   &      &&&\\
0 & \omega_0   & g\sqrt{2S} &         0 & \cdots                        &   &      &&&\\
0 & g\sqrt{2S} & \omega_{\rm s}   &         0 & \cdots                        &   &      &&&\\
0 & 0          & 0  & 2\omega_0 & g\sqrt{2S}\sqrt{2}       & 0 & \cdots&&&\\
0 & 0          & 0  & g\sqrt{2S}\sqrt{2}& \omega_0+\omega_{\rm s}& g\sqrt{4S-2} & \cdots&&&\\
0 & 0          & 0  & 0& g\sqrt{4S-2} & 2\omega_{\rm s}& 0 & \cdots&&\\
0 & 0          & 0 &0&0&    0 & n\omega_0 & gC_{S,-S}\sqrt{n}       & 0 & \cdots\\
0 & 0          & 0 &0&0&0 & gC_{S,-S}\sqrt{n}& \omega_0+\omega_{\rm s}& gC_{S,-S+1}\sqrt{n-1}  & \cdots\\
0 & 0          & 0 &0&0&0 & 0& gC_{S,-S+1}\sqrt{n-1} & 2\omega_{\rm s}&  \cdots\\
&&&& \cdots &&&&&
\end{array}\right)
\left(\begin{array}{c}
|0,-S\rangle \\
|1,-S\rangle \\ |0,-S+1\rangle \\
|2,-S\rangle \\ |1,-S+1\rangle \\ |0,-S+2\rangle \\
|3,-S\rangle \\ |2,-S+1\rangle \\ |1,-S+2\rangle \\ |0,-S+3\rangle \\
\ldots
\end{array}\right).
\normalsize
\eeq

We denote the eigenenergies in each sub-block characterized by $n_{\rm max}$ as
\beq
E(n_{\rm max},k), k=1,\cdots k_{\rm max}(n_{\rm max}),
\label{eigenE}
\eeq
where 
\beq
k_{\rm max}(n_{\rm max})=\left\{
\begin{array}{cll} 
n_{\rm max}+1 &
\quad {\rm for} \quad & n_{\rm max} \le N\\
N+1&
\quad {\rm for} \quad & n_{\rm max} > N.
\end{array}\right.
\eeq

We define the Rabi frequency as the energy difference (see \cite{Irinel} for analytical expression in the case $n_{\rm max}=1,2$ and 3): 
\beq
\omega_{\rm Rabi}(n_{\rm max},k)=E(n_{\rm max},k+1)-E(n_{\rm max},k), \quad
k=1, \cdots k_{\rm max}(n_{\rm max})-1.
\eeq

\begin{figure}
$$
\includegraphics{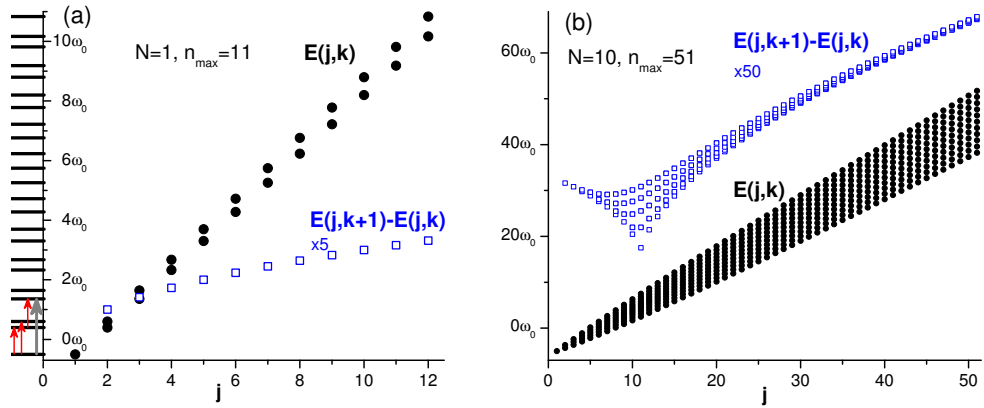}
$$
\caption{Eigen-energies (dots) and Rabi frequencies (blue squares) as a function of $n_{\rm max}$. (a) $N=1$. The bars denote the energy levels and the thin arrows denote one photon absorptions $(a^{\dagger})$. The bold arrow denotes an example of a nonlinear absorption $((a^{+})^2)$. The Rabi frequencies are multiplied by a factor of 5, for clarity.  (b) $N=10$: eigen-energies and Rabi frequencies (multiplied by 50 for clarity).}
\label{ENG}
\end{figure}

In Fig.~{\ref{ENG}(a)}, the eigenenergies for the case of $N=1$ (i.e., the Jaynes-Cummings model) are plotted by $\bullet$ as a function of the number of sub-block characterized by $n_{\rm max}$. The energies are also depicted by bars on the left side. The Rabi frequencies are plotted by squares (multiplied by 5). There are two states, $|n_{\rm max},m=-{1\over2}\rangle$ and $|n_{\rm max}-1,m={1\over2}\rangle$ in each subblock, and the difference of the two eigenstates (i.e. the Rabi frequency) is given by $\omega_{\rm Rabi}=2g\sqrt{n_{\rm max}}$. The bars denote the energy levels and the thin arrows denote one photon absorption processes $(a^{\dagger})$. The bold arrow denotes an example of 
a nonlinear absorption $((a^{\dagger})^2)$.  

We also give the photon number dependence of the energy and of the Rabi frequencies for the case $N=10$  in Fig.~\ref{ENG}(b). Here, the Rabi frequencies are multiplied by 50 for increased clarity. Note that the Rabi frequencies have a larger distribution when $N\simeq n$.

\subsection{Photon number dependence of the Rabi oscillation}

\begin{figure}
$$
\includegraphics{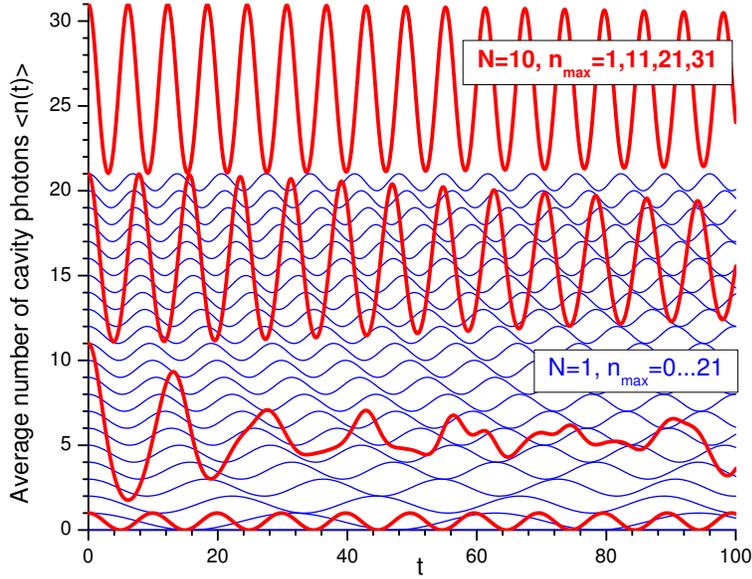}
$$
\caption{Rabi oscillations of the average number of photons, for various values of photon number. The thin blue curves show those for $N=1$ and $n_{\rm max}$ from 0 to 21 (starting from the bottom). The bold red curves are for $N=10$ and $n_{\rm max}=1,11,21$ and 31, starting from the bottom.}
\label{Rabi10}
\end{figure}

In Fig.~\ref{Rabi10}, we depict the Rabi oscillations versus time. The thin lines show the Rabi oscillations for the case of $N=1$, i.e., the case of Jaynes-Cummings model, where the Rabi oscillations in each sub-block is a sinusoidal curve with a single frequency $\omega=2g\sqrt{n_{\rm max}}$. The bold lines show the Rabi oscillations for the case of $N=10$. We find a simple oscillation for a small value ($n_{\rm max}=1)$, which corresponds to the vacuum-field Rabi splitting. We also find a simple oscillation for $n_{\rm max}=21$ in which the photon degree of freedom behaves as a classical field, and the Rabi frequencies takes values close to 
$\omega=2g|<a>|$. This oscillation corresponds to that of single spin in a driving field. If we see the graph more carefully, we find that the amplitude decreases slowly in the case of $n_{\rm max}=21$, which is due to a distribution of the frequencies. In the case of $n_{\rm photon}=31$, the oscillation is almost perfect sinusoidal within the observed time because the distribution is small. For larger values of $n_{\rm max}$ the oscillation looks perfect. Thus, we found two types of Rabi oscillations. The Rabi oscillation for small values of $n_{\rm max}$ corresponds to the cavity ringing\cite{Kaluzny_PRL83, Irinel}, while that for large $n_{\rm max}$ corresponds to the coherent Rabi oscillation which is found if the system is driven by an electromagnetic field\cite{Barbara,DeRaedt}.

Between them, at $n_{\rm photon}\sim N=10$, we find chaotic behavior which is attributed to the wide distribution of Rabi frequencies. This change between the two 
regions can be used to estimate how many spins contribute to the phenomena.

%===============================================================

\subsection{Dynamics in a dissipative environment} 

To study the dynamics of the system, we adopt a simple Lindblad type master equation 
of the density matrix of the system $\rho$ with damping terms\cite{Kubo,Louisell,Walls-Milburn}:
$$
{d\over dt}\rho={1\over i}\left[ {\cal H}, \rho\right]
-\kappa\left(
a^{\dagger}a\rho+\rho a^{\dagger}a-2a\rho a^{\dagger}
\right)
$$
\beq
-\gamma_{xy}\left(S^+S^{-}\rho+\rho S^+S^{-}-2S^{-}\rho S^{+} \right)
-\gamma_z\left((S^z)^2\rho+\rho (S^z)^2-2S^{z}\rho S^{z} \right).
\label{eqrho}
\eeq
The first term describes the coherent dynamics. In the next three terms the relaxation effect is introduced by a conventional Lindblad form. The second term shows the cavity decay rate $\kappa$, and the last two terms are giving the spin dephasing and relaxation, respectively. In general, there is an additional pumping term due to environment. This direct process could excite spins and thus generate emitted photons. But since the number of resonant bosons is generally low, we neglect these direct processes. Namely, we consider that the photon escapes from the system and is not created by the environment. We consider that the spin relaxation and dephasing is due only to energy relaxation processes, described here by constant values of $\kappa$, $\gamma_{xy}$, and $\gamma_z$.

For the Electron Spin Resonance (ESR) line shape in the case of linear response, we need an  distribution $\rho(0)$. The initial state, at the end of the pumping pulse, is an excited state. In order to express the degree of excitation, we used a steady-state distribution function at a temperature $T$:
\beq
\rho(0)={1\over Z}e^{-{\cal H}/k_{\rm B}T},
\quad Z={\rm Tr}e^{-{\cal H}/k_{\rm B}T}.
\eeq
Because the system is not in equilibrium, in this approximation the temperature $T$ can be regarded as a parameter describing the degree of freedom. A more extended treatment of the temperature effects has been formulated in the Uchiyama's paper\cite{uchiyama}, but here we take the present approximations.
%----------------------------------------------------------------------------------
\section{ESR spectrum}
The generalized susceptibility is given by the Kubo formular\cite{Kubo}, 

\beq
\chi(\omega)={\rm Tr}\int_{0}^{\infty} a^{\dagger}[a, \rho](t)e^{-i\omega t}dt
={\rm Tr}a^{\dagger}\rho_a[\omega].
\label{chiformula}
\eeq
%%%---------------------------------------------------
Here, the last term is derived as follows~\cite{uchiyama}.
Starting from the general form of the Kubo formula 
\beq
\chi_{BA}=\lim_{\varepsilon\rightarrow +0}{i\over\hbar}\int_0^{\infty}
dt e^{-i\omega t-\varepsilon t}{\rm Tr}[B(t),A]W,
\eeq
where $W$ is the stationary density matrix of the system, 
we use the following identities:
\beq
{\rm Tr}[B(t), A]W={\rm Tr} (B(t)AW-AB(t)W), \quad {\rm and}\quad B(t)=e^{iHt}B e^{-iHt}.
\eeq
Using trace properties regarding cyclic permutations, this leads to
\beq
{\rm Tr}[B(t), A]W = {\rm Tr} (B e^{-iHt}AW e^{iHt}-B e^{-iHt}WA e^{iHt})
={\rm Tr} (Be^{-iLt}[A,W]),
\eeq
where $e^{-iLt}X$ is the time evolution of operator $X$ $(\equiv e^{iHt}X e^{-iHt})$. 
Here we consider the driving force to be $\xi a^{\dagger}e^{-i\omega t}$ and analyze the response of the operator $a$. Therefore we have $A=a$, $B= a^{\dagger}$ and
$\rho_a[\omega]$ is the transformation of Tr $e^{-iLt}[A,W]$. With the except of the notation $[a, \rho(t)](t)$, Eq.~(\ref{chiformula}) is the standard Kubo formula for photon absorption.  In the same equation, we used the following notations:

%%%%--------------------------------------------------------
\beq
[a,\rho](t)=e^{-i{\cal L}t}[a,\rho(0)],
\eeq
where $ e^{-i{\cal L}t}$ denotes the time evolution of the density matrix (\ref{eqrho}) and 
\beq
\rho_a[\omega]=\int_{0}^{\infty}[a, \rho](t)e^{-i\omega t}dt.
\eeq

Here $a^{\dagger}$ is a time independent operator. The notation $[a,\rho](t)$ means $e^{-iHt} [a,\rho] e^{-iHt}$, where $a$ and $\rho$ are time independent operators. 
Thus $a^{\dagger}$ can be put out of the integral of Eq.~(\ref{chiformula}).

In the case of pure quantum dynamics, resonance frequencies of the linear response are given by
\beq
\omega_{\rm LR}(n_{\rm max},k,k')=E(n_{\rm max}+1,k')-E(n_{\rm max},k), 
\eeq
where
$$k=1,\cdots, k_{\rm max}(n_{\rm max}), k=1,\cdots, k_{\rm max}(n_{\rm max}+1), 
$$
at which the spectrum has delta-function peaks. The amplitude of each resonance is given by
\beq
\begin{array}{ccl}
I(\omega)&=&{\pi(e^{-\beta E(n_{\rm max},k)}-e^{-\beta E(n_{\rm max}+1,k')})\over Z}
|\langle n_{\rm max}+1,k'|a^{\dagger}| n_{\rm max},k\rangle|^2\\
&&
\delta(\omega-{E(n_{\rm max}+1,k')-E(n_{\rm max},k)\over \hbar}),\end{array}
\eeq
where $Z=\sum_{n_{\rm max}=0}\sum_k e^{-\beta E(n_{\rm max},k)}$.

\begin{figure}
$$
\includegraphics{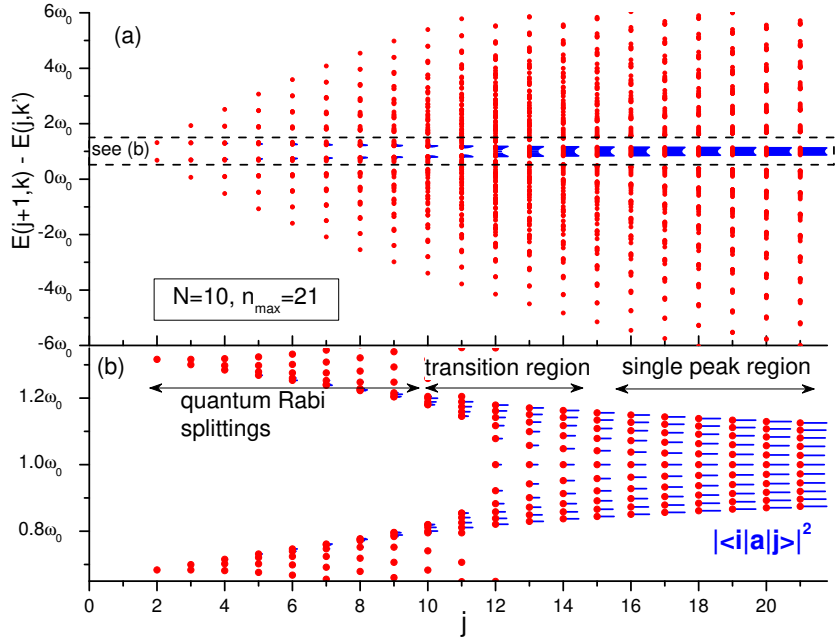}%
$$
\caption{(a) Resonance frequencies as a function of $n_{\rm max}$, for $N=10$. The bars denote the strength of the matrix elements of photon absorption for the resonance $|\langle i|a|j\rangle|^2$. (b) Zoom for frequencies around $\omega_0$. Only the resonances $\Delta E\sim \omega_0$ have visible transition probability. One distinguish three regions: the quantum region showing Rabi splittings, the transition region and the classical region where one observes one single ESR peak.
}
\label{Resonance50}
\end{figure}

We give the distribution of the linear response resonant frequency $\{\omega_{\rm LR}\}$ for the case with $N=10$ for each value of $n_{\rm max}$ in Fig.~\ref{Resonance50}. Because the interaction is described by $a^{\dagger}S^-+aS^+$, the states with $\Delta M=1$ (and thus $\Delta E\simeq \omega_0$) can be connected directly, while the other resonances are realized via higher order processes. Consequently, the transition with $\omega_{\rm LR}\simeq\omega_0$ has a large matrix element, while for the other resonances, the matrix elements are small and not visible in Fig.~\ref{Resonance50}.

In the large $n_{\rm max}$ limit, we may use the semi-classical picture for the electromagnetic field as $a\rightarrow |a|e^{i\omega t}$, and then the standard interaction Hamiltonian shows a rotating external field $|a|\left(S^+e^{i\omega t}+S^-e^{-i\omega t}\right)$. In this case, the resonance frequency is equal to $\omega_{\rm S}$, as expected for the classical ESR case. Thus Fig.~\ref{Resonance50} gives the transition from the quantum ringing (vacuum-field Rabi oscillation) region  to the classical field region as a function of the photon number.

In the dissipative environments, we adopt the quantum master equation (\ref{eqrho})
instead of the pure quantum dynamics. Because $[a, \rho](t)$ follows the equation of motion (\ref{eqrho}) \cite{uchiyama}, we use its Fourier-Laplace transformation:
\beq
i\omega\rho_a[\omega]-\rho_a(0)
={1\over i\hbar} \left[ {\cal H}_{\rm S}, \rho_a[\omega]\right]
-\kappa\left(
Z^{\dagger}Z\rho_a[\omega]-2Z\rho_a[\omega]Z^{\dagger}+\rho_a[\omega]Z^{\dagger}Z
\right),
\label{eqrhoFL}
\eeq
where $Z$ denotes $a$, $S^-$, or $S^z$ in the equation (\ref{eqrho}), with
\beq
\rho_a(0)=\left[a,\rho^{\rm eq}(T_{\rm system})\right],
\eeq
where $T_{\rm system}$ is the temperature of the system (cavity).

The equation (\ref{eqrhoFL}) is a set of linear equations of components of $\rho_a$ with the inhomogeneous term $\rho_a(0)$:
\beq
\rho_a(0)=i\omega\rho_a[\omega]
+i{1\over \hbar} \left[ {\cal H}_{\rm S}, \rho_a[\omega]\right]
+\kappa\left(
Z^{\dagger}Z\rho_a[\omega]-2Z\rho_a[\omega]Z^{\dagger}+\rho_a[\omega]Z^{\dagger}Z
\right).
\eeq
The explicit form of the equations for each matrix element $(ij)$ is given by
$$
\rho_a(0)_{ij}=i\omega\rho_a[\omega]_{ij}
+i{1\over \hbar} 
\sum_k\left(({\cal H}_{\rm S})_{ik}\rho_a[\omega]_{kj}
            -\rho_a[\omega]_{ik}({\cal H}_{\rm S})_{kj}
\right)$$
\beq
+\kappa\left(\sum_k\sum_n
Z^{\dagger}_{ik}Z_{kn}\rho_a[\omega]_{nj}
-2Z_{ik}\rho_a[\omega]_{kn}Z^{\dagger}_{nj}
+\rho_a[\omega]_{ik}Z^{\dagger}_{kn}Z_{nj}
\right).
\eeq

The ESR spectrum is obtained by the solution $\rho_a[\omega]$ of this equation 
for values of $\omega$ as
\beq
\chi(\omega)=\chi'(\omega)-i\chi''(\omega)={i\over\hbar}{\rm Tr}a^{\dagger}\rho_a[\omega].
\eeq

\begin{figure}
$$
\includegraphics{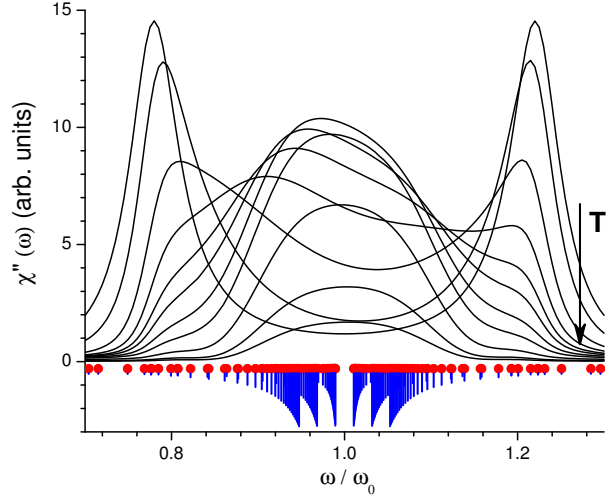}%
$$
\caption{Complex admittance for the case $N=5$ and $n_{\rm max}=21$, with $g=0.1, \kappa=0.01$, $\gamma_{xy}=0.01$ and $\gamma_z=0.01$. Here we plot data for $T=$0.1, 1, 2, 3, 4, 5, 7, 10, 20, 50, 100, in increasing order as shown by the arrow. The line shape changes from the double peak to the a single peak as the temperature increases. Below the axis, we plot the positions of the resonant frequencies (dots); the bar below each dot denotes the strength of the matrix element of the resonance, as in Fig.~\ref{Resonance50}.
}
\label{Chi5n10g01}
\end{figure}

In Fig.~\ref{Chi5n10g01}, we depict an example of temperature dependence of $\chi(\omega)$ for $N=5$, where the maximum value of $n_{\rm max}$ is 25. The parameters $g=0.1$, $\kappa=0.01,\gamma_{xy}=0.01$, and $\gamma_z=0.01$ are used. There we find that $\chi(\omega)$ has two peaks which correspond to the vacuum-field Rabi splitting\cite{Agarwal} at a low temperature ($T=0.01$). As the temperature increases, the resonances between excited states become to contribute and a single peak is formed.

%=====================================================================
\section{Non-equilibrium distribution of photons}
So far, we assumed that the photons in the cavity are in equilibrium at a certain temperature. But if we excite the cavity from outside by a finite driving force $\xi$, the distribution of photons can have a non-equilibrium form, as studied by Bishop et al.\cite{Bishop}. Also, nonlinear processes due to $(a^{\dagger})^k, k>1$ can contribute. The frequency of the nonlinear resonance is given by 
\beq
\left|\omega_{\rm NL}(n_{\rm max},k)={E(n,k)-E(n',k)\over n-n'}\right|, \quad
k=1,\cdots, k_{\rm max}(n_{\rm max}).
\eeq  

\subsection{Spectrum of a heterodyne measurement}

Under the driving force $\xi$, the system reaches a stationary state $\rho_{st}$
which is obtained as a stationary solution of (\ref{eqrho}).
However, the Hamiltonian (\ref{ham}) is time dependent and we cannot simply put $d\rho/dt=0$. Thus, we use a rotating frame to remove the time dependence $e^{\pm i\omega t}$:
\beq
{\cal H}'=U{\cal H}U^{-1}, \quad U=\exp \left(i\omega t (a^{\dagger}a+S^z)\right).
\eeq
This transforms quantities as
\beq
Ua^{\dagger}U^{-1}=a^{\dagger}e^{i\omega t}, \quad US^{-}U^{-1}=S^{-}e^{-i\omega t}
\eeq
and 
$$
{dU\rho U^{-1}\over dt}=
U{d\rho\over dt} U^{-1}+{dU\over dt}\rho U^{-1}+dU\rho {dU^{-1}\over dt}
$$
\beq
=U{d\rho\over dt} U^{-1}
+i\omega \left( (a^{\dagger}a+S^z)U\rho U^{-1}-U\rho U^{-1}(a^{\dagger}a+S^z)\right).
\eeq
Therefore, with this change, the equation of motion for $\rho'=U\rho U^{-1}$ is given by  (\ref{eqrho}) with the following form of the Hamiltonian:
\beq
{\cal H}'=\omega_0a^{\dagger}a+\omega_{\rm s}(\sum_i^N S_i^z + N/2) 
-g\left(a^{\dagger}\sum_i^NS_i^- + a\sum_i^NS_i^+ \right)
+\xi\left(a^{\dagger} + a \right)
-\omega (a^{\dagger}a+S^z).
\label{hamrot}
\eeq 

Now, we consider the dynamics in this frame, and obtain a stationary state by setting $d\rho/dt=0$:
$$
{1\over i\hbar}\left[ {\cal H}', \rho_{\rm st}\right]
-\kappa\left(
a^{\dagger}a\rho_{\rm st}+\rho_{\rm st} a^{\dagger}a-2a\rho_{\rm st} a^{\dagger}
\right)
$$
\beq
-\gamma\left(S^+S^{-}\rho_{\rm st}+\rho_{\rm st} S^+S^{-}-2S^{-}\rho_{\rm st}S^{+} \right)
-\gamma_z\left((S^z)^2\rho_{\rm st}+\rho_{\rm st} (S^z)^2-2S^{z}\rho_{\rm st}S^{z} \right)=0.
%\label{eqrho}
\eeq
This is a homogeneous equation of $\rho_{\rm st}$\cite{Bishop}. The expectation values
\beq
I={\rm Tr} (a+a^{\dagger})\rho_{\rm st}, \quad {\rm and } \quad 
Q={\rm Tr} (ia^{\dagger}-ia)\rho_{\rm st}
\eeq
correspond to the heterodyne measurement of the field quadratures. The steady-state transmission amplitude is given by $Q$. For a small strength $\xi=0.001$, the transmission shows a peak corresponding to the vacuum-field Rabi splitting. 

\begin{figure}
$$
\includegraphics{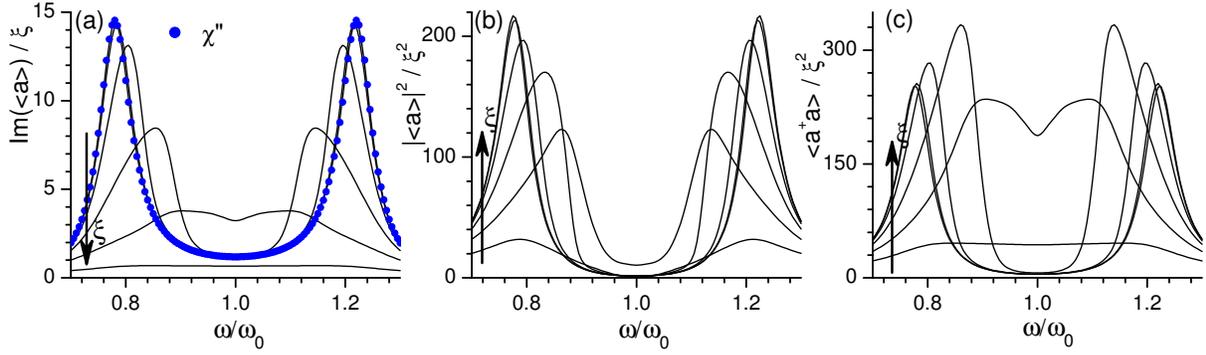}
$$
\caption{(a) Dependence of $Q/2\xi={\rm Im}(<a>)/\xi$ on the driving frequency $\omega$, for different $\xi$. The blue dots show the linear response case (see Fig. 4, T=0.1) (b) Driving frequency $\omega$ dependence of $|\langle a\rangle|^2/\xi^2$ of stationary states under various strengths of the driving force $\xi$ (c)Driving frequency $\omega$ dependence of the photon number of stationary states $<a^{\dagger}a>/\xi^2$ under various strengths of the driving force $\xi$. The parameters are $N=5$ and $n_{\rm max}=25$, with $g=0.1, \kappa=0.01$ and $\gamma_{xy}=\gamma_z=0.01$. The strengths of $\xi$ are 0.01,0.02,0.05,0.1,0.2, and 0.5, in increasing order as indicated by the arrows in each figure.}
\label{DRVa}
\end{figure}

In Fig.~\ref{DRVa}(a), we compare the line shape of the linear response $\chi(\omega)$ ($\circ$), and that of $Q/2\xi$ for various values of $\xi$. We find that the spectrum of $Q$ for small values of $\xi$ agree with $\chi(\omega)$. We plot the spectrum of $|\langle a\rangle|^2/\xi^2$ in Fig.~\ref{DRVa}(b). Here we find that $\chi(\omega)$ tends to move to a single peak, but $|\langle a\rangle|^2$ keeps a double-peak structure. We also show the spectrum of $\langle a^{\dagger}a\rangle/\xi^2$ in Fig.~\ref{DRVa}(c). In cases with many spins and photons, there exist many resonant modes within the vacuum-field Rabi splitting, and the double peak structure is affected. 

In strong drive, the photon number shows a single peak, while the heterodyne signal has a double-peak shape. This can be understood from a view point of the phase decoherence of the photon mode.

In the case $\omega=\omega_0$,

$$
|\psi(t)\rangle_{\rm ph}=e^{i\xi(a^{\dagger}+a)t}|\psi(0)\rangle_{\rm ph}
= e^{-{|\xi t|^2\over2}}\sum_{n=0}^{\infty}{(i\xi t)^n(a^{\dagger})^n\over n!}|0\rangle
$$
\beq
=e^{-{|\xi t|^2\over2}}\sum_{n=0}^{\infty}{(i\xi t)^n\over\sqrt{n!}}|n\rangle,
\eeq
where we used the relation
\beq
|n\rangle={1\over \sqrt{n!}}(a^{\dagger})^n|0\rangle.
\eeq
In the case with $|\psi(0)\rangle_{\rm ph}=|0\rangle$, the number of photons increases with time:
\beq
a|i\xi t\rangle = i\xi t|i\xi t\rangle  \quad {\rm and} \quad
\langle n\rangle \propto (\xi t)^2.
\eeq

The damping term causes a saturated value of the photon number in the stationary state. The dissipation term causes phase decoherence of the coherent state, due to the phase fluctuation 
\beq 
a|i\xi t\rangle = e^{i\delta} (i\xi t)|i\xi t\rangle. 
\eeq
By averaging over $\delta$, the $|\langle a\rangle|$ is strongly reduced, while on the other hand, $\langle a^{\dagger}a\rangle$ is robust against this effect. This explains the large reduction of the spectrum at its center $\omega\simeq \omega_0=\omega_{\rm S}$, where the cavity photons are incoherently excited.

%%===============================================================================
\section{Dissipation of the total spin}

The total spin is given by
\beq
S(S+1)=M_x^2+M_y^2+M_z^2.
\eeq
So far, we studied the case where the total spin is fixed to be the maximum value
($S=N/2$). This value is for the state with all the spin aligned. When the system is excited, the value of the magnetization $M=-S+$({\rm number of up spins}) increases. 
In order to conserve the total spin, the $xy$ components must be coherent. In the model Hamiltonian (\ref{ham}), the total spin is conserved and the dissipation dynamics (\ref{eqrho}) does not affect the total spin, either. Therefore, in the present model, when the spins are excited, namely they are polarized longitudinally,
the coherence of the transverse component appears. This large coherent transverse component would cause the super-radiance\cite{Dicke}.

However, in practice, we may expect the decoherence of the spin state. In this case, the total spin is reduced. While the maximum of $\mbold{S}^2$ is given by $S(S+1)=N(N+2)/4$, the actual value can be reduced to the value corresponding to $N$ phase-independent spins
\beq
\langle \phi|\mbold{S}^2|\phi\rangle={3N\over4},
\eeq
where
\beq
|\phi\rangle={1\over 2^{N/2}}\sum_{\{\sigma_i=\pm 1\}} 
|\sigma_1,\sigma_2,\cdots,\sigma_N\rangle.
\eeq
This implies that we cannot use the Tavis-Cummings model with only one value of $S$.
The effective value of $S$ may change for each frequency. Thus we must adopt a Hamiltonian with $2^N$ dimensional spin space instead of $2S+1$.

\begin{figure}
\includegraphics{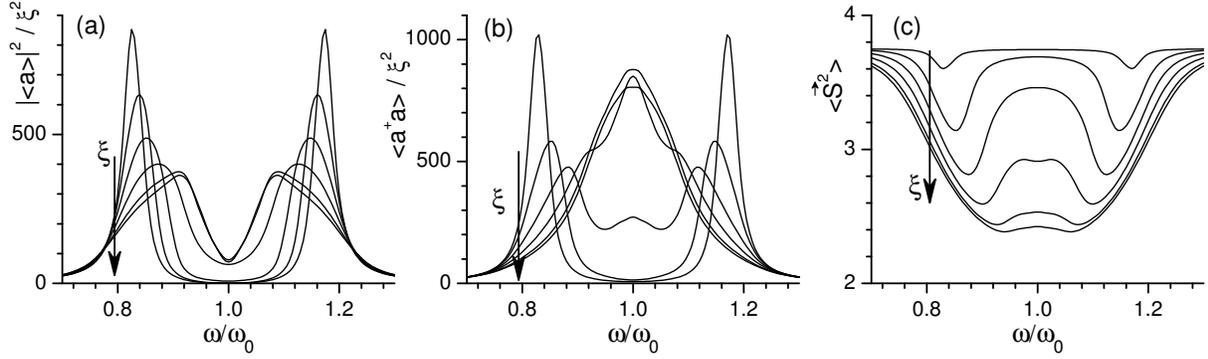}%
\caption{Frequency dependences of 
(a) $|\langle a\rangle|^2/\xi^2$,
(b) $\langle a^{\dagger}a\rangle/\xi^2$, and 
(c)$|\langle S^2\rangle$ as a function of the
frequency of the driving force with the dissipation term Eq.~(\ref{eqrho2}) for $N=3,n_{\rm max}=20$ with $g=0.1, \kappa=0.01$ and $\gamma_{xy}=\gamma_z=0.01$
The values of $\xi$ are 0.01, 0.03, 0.05, 0.07, 0.09 and 0.1, in increasing order as shown by an arrow in each figure. 
}
\label{FULLC3}
\end{figure}

To take into account this fact, we adopt a new type of dissipation mechanism, 
in which the total spin can change. In the process given by the master equation (\ref{eqrho}), the environment couple to the total magnetization $M_z=\sum_{i=1}^NS^z_i$, and the dissipation effect was given by the last term in (\ref{eqrho}). In contrast, now we provide the thermal contact to each spin, and we have a new master equation:
$$
{d\over dt}\rho={1\over i\hbar}\left[ {\cal H}, \rho\right]
-\kappa\left(
a^{\dagger}a\rho+\rho a^{\dagger}a-2a\rho a^{\dagger}
\right)
$$
\beq
-\gamma_{xy}\sum_{i=1}^N\left(S_i^+S_i^{-}\rho+\rho S_i^+S_i^{-}-2S_i^{-}\rho S_i^{+} \right)
-\gamma_z\sum_{i=1}^N\left((S_i^z)^2\rho+\rho (S_i^z)^2-2S_i^{z}\rho S_i^{z} \right).
\label{eqrho2}
\eeq
Using the present master equation in the extended space, we present in Fig.~\ref{FULLC3} the frequency-dependences of $|\langle a\rangle|^2$, $\langle a^{\dagger}a\rangle$ (normalized to $\xi^2$), and $|\langle 2M+N\rangle|^2$ for stationary states. For $N=3$, the maximum value of $\mbold{S}^2$ is $S(S+1)= 3.75$, and its minimum value is $3N/4=2.25$~\cite{morespins}. Here we find that the total spin is largely reduced, and almost reaches the minimum value.

Thus, we expect that even if the number of spins is large in the cavity, practical values of $S$ could be small. In the thermodynamic equilibrium state, we must estimate how many number of spins in the material contribute coherently (see next Section).

\section{Experimental study}

\subsection{Strong coupling regime}
We have carried out a spectroscopic study of a large $N=N_{spins}$ sample, in the strong coupling regime. The system of choice is the two-level spin system of the dipheriyl-picri-hydrazyl (DPPH), a well known Electronic Spin Resonance (ESR) standard ~\cite{Krzystek_JMR97}. Each DPPH molecule contains a single spin $S=1/2$ and therefore the spins are relatively diluted and their environment is lacking anisotropy ($g_S\simeq 2$). As a consequence, the line width of the DPPH resonance is very narrow, of the order of 0.1 mT. It is only at very high Zeeman splitting (large static fields and large frequencies, of few hundreds GHz) that the anisotropy in the g-factor starts to be detectable, leading to an increased resonance linewidth. But at frequencies $\sim 10$~GHz as in our study, DPPH is among the best candidates to study resonance splitting consequent to cavity photons rather than anisotropy in spin environment. 

The sample is placed inside a cylindrical cavity operated in the TE$_{011}$ mode, with an optimized coupling to the microwave B-field. The resonance mode is located at $\omega_0/2\pi=9.624$~GHz and the system is operated in reflection: a microwave pulse is pumped into the cavity and the reflected pulse is detected by a home build heterodyne setup. More precisely, the reflected pulse is down-mixed by another pulse with a frequency close to the one used in the pump pulse (in our case, the shift is of 100~MHz and the pump pulse has a length of 1~$\mu$s).

\begin{figure}
$$
\includegraphics{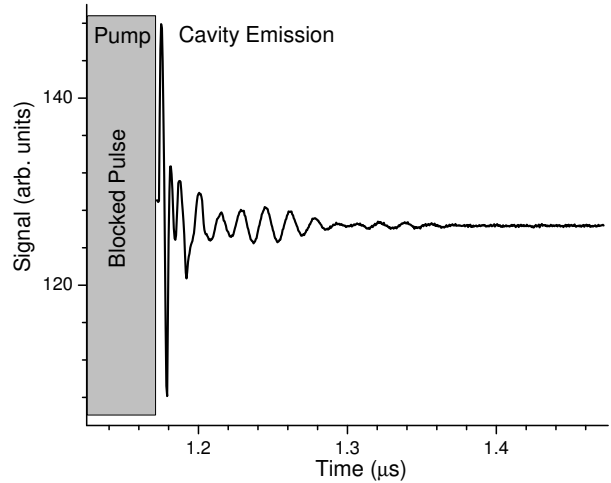}
$$
\caption{Example of cavity ringing detected following a pump pulse. The readout is blocked during pumping, to avoid potentially large signals into the amplification stage. The cavity emission shows beatings which are visible as distinct peaks in the FFT trace (see Fig.~\ref{figexp}).}
 \label{figexp2}
\end{figure}

To study the eigenenergies of the coupled spin-photon system, we use the microwave pulse at $\omega$ to excite the system and record the photon release after the excitation. When the microwave is switched off, the cavity coherently emits photons corresponding to its own eigenmodes. The time evolution of such a signal is shown in Fig.~\ref{figexp2} where one can clearly see beatings of close but distinct frequencies in the emitted signal. This technique is called cavity ringing and allows to study the energy spectrum by detecting photon frequencies in the Fast Fourier Transform (FFT) of decaying ringing oscillations. Obviously, most of the FFT spectrum is localized at the pump frequency. Therefore, we detune slightly the pump from the cavity resonance (by 50 MHz), in order to see the cavity-spin eigenmodes in the tail of the FFT.

\begin{figure}
$$
\includegraphics{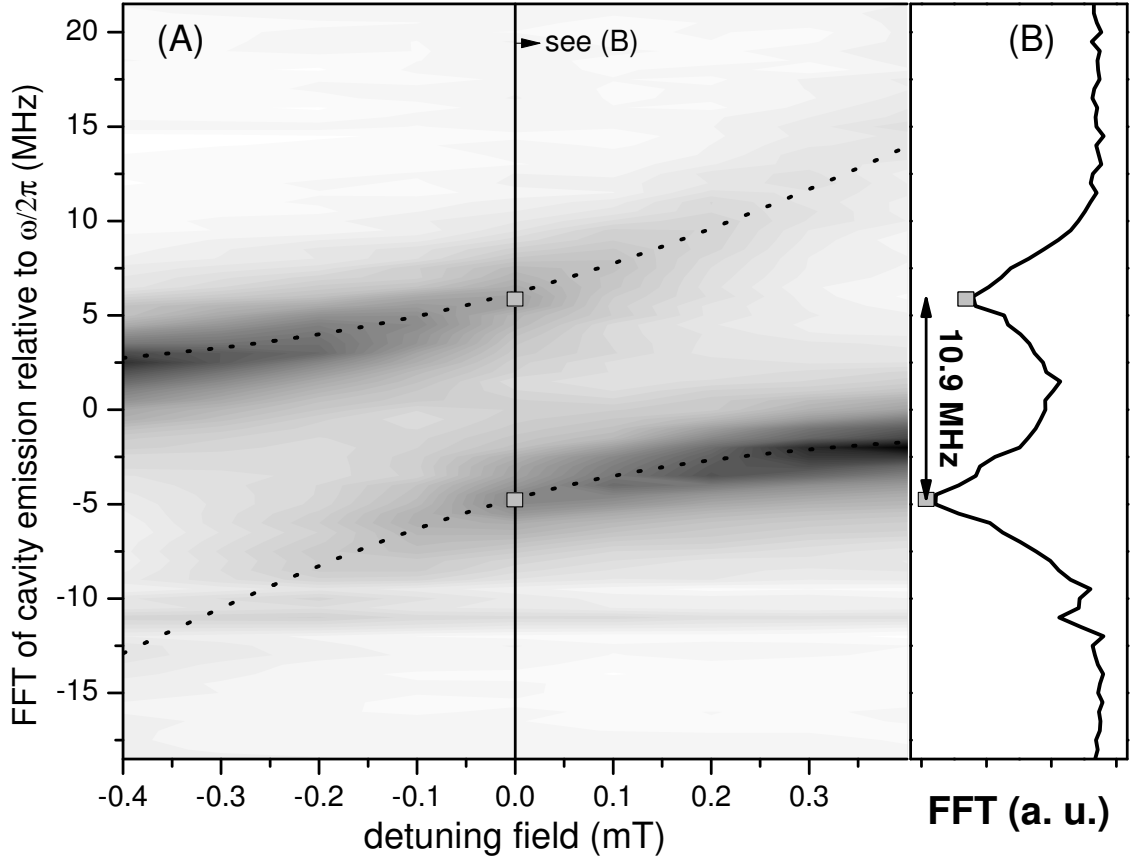}
$$
\caption{{\rm (a)} Intensity plot of the Fourier transform of detected cavity ringing, measured at room temperature. The plot is given as a function of the magnetic field around the resonance condition (FFT frequencies are relative to the cavity resonance). One observes a level repulsion developing in the resonance window, due to the spin-photon strong coupling. The effect is particularly visible at zero detuning (vertical marking line). {\rm (b)} Actual FFT trace recorded at zero detuning showing a measured Rabi splitting of 10.9~MHz. The peaks are distinguishable by a splitting larger than peaks width. } \label{figexp}
\end{figure}

Measured Fourier traces as a function of the applied static field are presented in Fig.~\ref{figexp}. The FFT intensity is gray coded, the dark areas representing the eigenmodes of the spin-photon coupled system. Frequencies are relative to the sample-loaded cavity resonance, in absence of applied field. The static field is varied around the empty cavity resonance condition $\mu_0H_z=\hbar\omega_0/g_s\mu_B$, with $g_s=2$ by a detuning is given by $\mu_0\delta H_z=-\hbar\Delta/g_s\mu_B$. At zero detuning, the actual FFT plot is given in Fig.~\ref{figexp} and one clearly distinguish (that is, peaks separation is larger than peak width) two eigenenergies split by 10.9 MHz. The continuous lines are fits made considering a level repulsion $\Delta$.

In our experimental conditions, due to the spin-photon strong coupling, one observes two peaks separated by the characteristic Rabi splitting. The strength of the coupling can be estimated using the model of $N$ spins coupled to a single photon, although the cavity contains a large number of $n_{\rm max}$ photons. As we see in Fig.~\ref{Resonance50}, as long as the number of photons ($n_{\rm max}$) is much smaller than the number of spins $N$, the resonance frequencies are given by  
$E(1,k)-E(0,1)$, $k=1,2$ of Eqs.~(\ref{eigenE}). Thus, the difference of the resonance frequencies is given by that of the vacuum-field Rabi oscillation. Therefore, the cavity ringing shows coherent oscillations with the frequencies $\omega_{e0,1}/2\pi$, given by:
\beq
\omega_{e0,1}-\omega=-{\Delta\over2}\pm{1\over2}\sqrt{\Delta^2+\Omega_R^2},
\label{em}
\eeq
on which the continuous lines of Fig.~\ref{figexp}(a) are based. The fit procedure leads to a Rabi splitting of $\Omega_R/2\pi=10.9$~MHz.

\subsection{Photon number and the types of Rabi oscillation}

As mentioned previously, it is important to compare the number of spins (or total spin size) to the number of photons, since quantum peaks require large splitting (large $N$) and a smaller number of photons. In our experiment, taking into account the DPPH spin density, we estimate the number of spins to be on the order of $10^{20}$. Since the temperature is $T=300K$, the size of the total spin will be reduced due to the Boltzmann population of the two levels . For a Zeeman splitting of $\omega=9.62$ GHz, the magnetization is estimated as 
\beq
m(T=300K)={N_{\rm spins}\over2}\tanh\left({\hbar\omega\over2 k_{\rm B}T}\right)\simeq 10^{-3}{N_{\rm spins}\over2},
\eeq 
and thus the dominant value of the total spin is of the order
\beq
S = m(T=300K) \sim 10^{-3}{N_{\rm spins}\over2}.
\eeq
The number of photons inside the cavity can be roughly estimated by using the $Q$ factor of the cavity ($Q\sim 1000$) to get a photon life time of less than 100 ns and estimations of the microwave power delivered by the setup. We find a range for the number of photons of the order of $10^{12}-10^{14}$ which fulfills the condition:
\beq
n_{\rm photon} \ll S
\eeq
In this case, we expect indeed that the vacuum-field Rabi splitting of the resonance  frequency to be clearly found. The splitting is measured to be of the order $10$MHz:
\beq
\Omega_{\rm R}/2\pi \simeq g\sqrt{S}=10{\rm MHz},
\eeq
from which we can estimate an order of magnitude for the coupling constant $g\sim 10^{-2} - 10^{-1}$ Hz.

On the other hand, recent experiments demonstrate the existence of coherent Rabi oscillation of diluted spin systems driven by an electromagnetic field \cite{Barbara, DeRaedt}. In those experiments the number of photons in the cavity is much larger than $N$ as we see in Fig.~\ref{Rabi10}. It is therefore an interesting future problem to study the transition between these two regions (i.e., the region of the vacuum-field Rabi oscillation ($n\ll N$) and that of the field driven Rabi oscillation ($n \gg N$).

\section{Summary and discussion}

We studied the line shape as a function of the size of an ensemble of spins and the cavity mode photon number. We found the line shapes change depending on the number of spins, number of photon, and also types of dissipation. 

We found two regions of the photon numbers. One of them is the cavity ringing region
where $N\gg n$, while the other is that of coherent Rabi oscillation under driving
electromagnetic field.

We presented an experimental data of the hybridization of the cavity and the spins
which shows double peaks in the spectrum representing the vacuum-field Rabi oscillation due to the cavity ringing, and we estimate the number of spins and photons contributing to the coherent dynamics.

With our model, the role of other factors in the spread of Rabi splitting, such as: dipolar or hyperfine interactions, local anisotropic crystal fields, and the size of the sample vs. size of a cavity (mode) can be studied as well. It is also an interesting problem to study the relation between this dissipation effect and the existence of the optical bistability\cite{opticalbistability,Drummond,Gripp,Shirai}. If the dissipation of the transverse coherence does not exist, the polarization of the system is monotonic with the driving force. But, the dissipation can cause a nonlinear dependence which leads to optical bistability.

\section*{Acknowledgments}
This work was partially supported by the Mitsubishi foundation, 
the NSF Cooperative Agreement Grant No. DMR-0654118 and No. NHMFL-UCGP 5059, NSF grant No. DMR-0645408, National Computer Facilities (Netherlands) and also the Next Generation Super Computer Project, Nanoscience Program of MEXT. 
The computation in this work has been done using the facilities of the Supercomputer Center, Institute for Solid State Physics, University of Tokyo.
%Numerical calculations were done on the supercomputer of ISSP. 
We thank City of Marseille and Aix Marseille University for financial support.

%====================================================================


\section*{References}
\begin{thebibliography}{99}

\bibitem{JC}  Jaynes E. T. and  Cummings F. W. 1963, \textit{Proc. IEEE} {\bf 51}, 89.

\bibitem{TC} Tavis M.  and Cummings  F. W., 1968, \textit{Phys. Rev.} {\bf 170}, 379.

\bibitem{Kaluzny_PRL83}  Kaluzny Y.,  Goy P.,  Gross M., Raimond J. M.,  Haroche S.,
1983 \textit{Phys. Rev. Lett.} {\bf 51}, 1175-1178.

\bibitem{Mondragon_PRL83}  Sanchez Mondragon J. J., Narozhny N. B. , and Eberly  J. H. , 1983 \textit{Phys. Rev. Lett.} {\bf 51}, 550.

\bibitem{Thompson_PRL92} Thompson R. J., Rempe G., Kimble H. J., 1992, \textit{Phys. Rev. Lett.} {\bf 68}, 1132.

\bibitem{Brune_PRL96} Brune M., Schmidt-Kaler F., Maali A., Dreyer J., Hagley E., Raimond J. M., and Haroche S., 1996 \textit{Phys. Rev. Lett.} {\bf 76}, 1800.

\bibitem{Rabl_PRL06} Rabl P., et al., 2006, \textit{Phys. Rev. Lett.} {\bf 97}, 033003.

\bibitem{Imamoglu_PRL09} Imamo\u{g}lu A.,2009, \textit{Phys. Rev. Lett.} {\bf 102}, 083602.

\bibitem{Irinel} Chiorescu I., Groll N., Bertaina S., Mori T., and Miyashita S.,
2010, \textit{Phys. Rev. B} {\bf 82}, 024413.

\bibitem{Schuster} Schuster D. I., Sears A. P., Ginossar E., DiCarlo L.,
 Frunzio L., Morton J. J. L., Wu H., Briggs G. A. D., Buckley B. B.,
 Awschalom D. D., and Schoelkopf R. J., 2010 \textit{Phys. Rev. Lett.} {\bf 105}, 140501.

\bibitem{KuboY} Kubo Y.,. Ong F. R, Bertet P., Vion D., Jacques V., Zheng D.,
 Dreau A., Roch J.-F., Auffeves A., Jelezko F., Wrachrup J., Barthe M.F.,
 Bergonzo P., and Esteve D., 2010, \textit{Phys. Rev. Lett.} {\bf 105},  140502.

\bibitem{Amsuss_PRL11} Ams\"{u}ss R. et al., 2011, \textit{Phys. Rev. Lett.} {\bf 107}, 060502.

\bibitem{Semba_Nature11} Zhu X., Saito S., Kemp A., Kakuyanagi K.,
 Karimoto S., Nakano H., Munro W. J., Tokura Y., Everitt M. S., Nemoto K.,
 Kasu M., Mizuochi N., Semba K., 2011, \textit{Nature} {\bf 478}, 221.

\bibitem{KuboY_PRL11} Kubo Y., Grezes C., Dewes A., Umeda T., Isoya J., Sumiya H.,
 Morishita N., Abe H., Onoda S., Ohshima T., Jacques V., Dreau A., Roch J.-F.,
 Diniz I., Auffeves A., Vion D., Esteve D., Bertet P., 2011, \textit{Phys. Rev. Lett.} {\bf 107}, 220501.

\bibitem{Agarwal} Agarwal  G. S., 1984,  Phys. Rev. Lett. {\bf 53}, 1732. 
 Houdre R., Stanley R. P., and Ilegems M., 1996, Phys. Rev. A{\bf 53} 2711.
 ZhuY., Gauthier  D. J., Morin S. E., Wu Q., Carmichael H. J., and Mossberg T. W., 
1990, \textit{Phys. Rev. Lett.} {\bf 64}, 2499.

\bibitem{Bishop} Bishop L.S., Chow J. M., Koch J., Houck A. A., Devoret M. H.,
 Thuneberg E., Girvin S. M., Schoelkopf R. J., 2009, \textit{Nature Physics} {\bf 5}, 105.

\bibitem{Barbara} Bertaina S., Gambarelli S., Tkachuk A., Kurkini N., Malkin B.,
 Stepanov A., and Barbara B., 2007, \textit{Nat. Nanotechnol.} {\bf 2}, 39.
 Bertaina S., Gambarelli S., Mitra T., Tsukerblat B., Muller A., and Barbara B., 2008, \textit{Nature} (London) {\bf 453}, 203.
 Bertaina S., Chen  L., Groll N., Van Tol  J., Dalal N. S., and Chiorescu I., 2009, \textit{Phys. Rev. Lett.} {\bf 102}, 050501.
Bertaina S., Shim J. H., Gambarelli S., Malkin  B. Z., and Barbara B., 2009, 
\textit{Phys. Rev. Lett.} {\bf 103}, 226402.
 Bertaina S., Groll N., Chen L., and Chiorescu I., 2011, \textit{Phys. Rev. B} {\bf 84}, 134433.

\bibitem{DeRaedt}De Raedt H., Barbara B, Miyashita S.,  Michielsen K., Bertaina S, and Gambarelli S, 2012, \textit{Phys. Rev. B }{\bf 85}, 014408.

\bibitem{Kubo} Kubo R., Toda M. and Hashitsume N., {\it Statistical Physics II}, (Springer-Verlag, New York, 1085).

\bibitem{Louisell} Lousell W. H., {\it Quantum Statistical Properties of Radiation}, (Wiley, New York, 1973).

\bibitem{Walls-Milburn} Walls  D. F. and Milburn G. J., \textit{Quantum Optics} (Springer-Verlag, 1995).

\bibitem{uchiyama} Uchiyama C., Aihara M., Saeki M. and Miyashita S., 2009, \textit{Phys. Rev. E}, 80, 021128 1-16.

\bibitem{Dicke} Dicke R. H., 1954, Phys. Rev. {\bf 93}, 99.

\bibitem{morespins}We are preparing the massive calculation needed for a larger number of spins.

\bibitem{Krzystek_JMR97} Krzystek J., Sienkiewicz A., Pardi L., Brunel L.C., 1997,
 \textit{J. Magn. Reson.} {\bf 125(1)}, 207.


\bibitem{opticalbistability} Lugiato L.A., in {\it Progress in Optics} Vol. XXI, ed by E. Wolf pp. 69-216.

\bibitem{Drummond} Drummond P. D., 1981, IEEE, \textit{J. Quantum Electronics}, QE-{\bf 17}, 301.

\bibitem{Gripp} Gripp J., Mielke S. L., and Orozco L. A., and Carmichael H. J., 1996,
\textit{Phys. Rev. A} {\bf 54}, R3746.
 Raizen M. G., Thompson R. J., Brecha R. J., Kimble H. J., and Carmichael H. J.,
1989, \textit{Phys. Rev. Lett.} {\bf 63} 240.

\bibitem{Shirai} Shirai T., Mori T. and Miyashita S., unpublished.

\end{thebibliography}
\end{document}